\documentclass[10pt,a4paper,reqno]{amsart}
\usepackage{acronym}
\usepackage{amsfonts}
\usepackage{amsmath}
\usepackage{amssymb}
\usepackage{amsthm}
\usepackage{bm}
\usepackage{braket}
\usepackage{cite}
\usepackage{color}
\usepackage{geometry}
\usepackage{graphicx}
\usepackage{hyperref}
\usepackage{mathrsfs}
\usepackage{mathtools}
\usepackage{setspace}
\usepackage{stmaryrd}
\usepackage{subcaption}
\usepackage{tikz}

\newgeometry{top=1.5cm,bottom=1.5cm,outer=1.5cm,inner=1.5cm}

\newcommand{\bse}{\begin{subequations}}
\newcommand{\ese}{\end{subequations}}

\numberwithin{equation}{section}

\DeclareMathOperator{\tr}{tr}

\def\undertilde#1{\mathord{\vtop{\ialign{##\crcr
$\hfil\displaystyle{#1}\hfil$\crcr\noalign{\kern1.5pt\nointerlineskip}
$\hfil\widetilde{}\hfil$\crcr\noalign{\kern-6.5pt}}}}}
\def\underhat#1{\mathord{\vtop{\ialign{##\crcr
$\hfil\displaystyle{#1}\hfil$\crcr\noalign{\kern1.5pt\nointerlineskip}
$\hfil\widehat{}\hfil$\crcr\noalign{\kern-6.5pt}}}}}
\def\underbar#1{\mathord{\vtop{\ialign{##\crcr
$\hfil\displaystyle{#1}\hfil$\crcr\noalign{\kern1.5pt\nointerlineskip}
$\hfil\bar{}\hfil$\crcr\noalign{\kern-6.5pt}}}}}

\newcommand{\rT}{\mathrm{T}}
\newcommand{\rD}{\mathrm{D}}
\newcommand{\rd}{\mathrm{d}}

\newcommand{\bLd}{\mathbf{\Lambda}}
\newcommand{\tbLd}{{}^{t\!}\boldsymbol{\Lambda}}

\newcommand{\bO}{\boldsymbol{O}}

\newcommand{\bOa}{\boldsymbol{\Omega}}
\newcommand{\tbOa}{{}^{t\!}\boldsymbol{\Omega}}
\newcommand{\bC}{\boldsymbol{C}}
\newcommand{\tbC}{{}^{t\!}\boldsymbol{C}}
\newcommand{\bU}{\boldsymbol{U}}
\newcommand{\tbU}{{}^{t\!}\boldsymbol{U}}
\newcommand{\bu}{\boldsymbol{u}}

\newcommand{\bc}{\boldsymbol{c}}
\newcommand{\tbc}{{}^{t\!}\boldsymbol{c}}
\newcommand{\bA}{\boldsymbol{A}}

\newcommand{\bM}{\boldsymbol{M}}

\newcommand{\bV}{\boldsymbol{V}}

\newcommand{\Oa}{\Omega}

\acrodef{2D}[2D]{two-dimensional}
\acrodef{2DTL}[2DTL]{two-dimensional Toda lattice}
\acrodef{3D}[3D]{three-dimensional}
\acrodef{ABS}[ABS]{Adler--Bobenko--Suris}
\acrodef{bDT}[bDT]{binary Darboux transform}
\acrodef{BT}[BT]{B\"acklund transform}
\acrodef{BSQ}[BSQ]{Boussinesq}
\acrodef{CAC}[CAC]{consistency-around-the-cube}
\acrodef{DT}[DT]{Darboux transform}
\acrodef{DL}[DL]{direct linearisation}
\acrodef{DLT}[DLT]{direct linearising transform}
\acrodef{DDeltaE}[D$\Delta$E]{differential-difference equation}
\acrodef{FX}[FX]{Fordy--Xenitidis}
\acrodef{GD}[GD]{Gel'fand--Dikii}
\acrodef{KP}[KP]{Kadomtsev--Petviashvili}
\acrodef{KdV}[KdV]{Korteweg--de Vries}
\acrodef{HM}[HM]{Hirota--Miwa}
\acrodef{MDC}[MDC]{multi-dimensional consistency}
\acrodef{NQC}[NQC]{Nijhoff--Quispel--Capel}
\acrodef{ODE}[ODE]{ordinary differential equation}
\acrodef{ODeltaE}[O$\Delta$E]{ordinary difference equation}
\acrodef{PDE}[PDE]{partial differential equation}
\acrodef{PDeltaE}[P$\Delta$E]{partial difference equation}
\acrodef{RHP}[RHP]{Riemann--Hilbert problem}
\acrodef{sG}[sG]{sine--Gordon}
\acrodef{YB}[YB]{Yang--Baxter}

\title[On nonautonomous differential-difference AKP, BKP and CKP equations]{On nonautonomous differential-difference AKP, BKP and CKP equations}
\author{Wei Fu}
\address[WF]{School of Mathematical Sciences and Shanghai Key Laboratory of Pure Mathematics and Mathematical Practice \\
East China Normal University \\ 500 Dongchuan Road \\ Shanghai 200241 \\ People's Republic of China}
\author{Frank W. Nijhoff}
\address[FWN]{School of Mathematics \\ University of Leeds \\ Leeds LS2 9JT \\ United Kingdom}

\begin{document}

\begin{abstract}
Based on the direct linearisation framework of the discrete Kadomtsev--Petviashvili-type equations presented in [Proc. R. Soc. A, \textbf{473} (2017) 20160915],
six novel nonautonomous differential-difference equations are established, including three in the AKP class, two in the BKP class and one in the CKP class.
In particular, one in the BKP class and the one in the CKP class are both in (2+2)-dimensional form.
All the six models are integrable in the sense of having the same linear integral equation representations as
those of their associated discrete Kadomtsev--Petviashvili-type equations,
which guarantees the existence of soliton-type solutions and the multi-dimensional consistency of these new equations
from the viewpoint of the direct linearisation.
\end{abstract}


\maketitle

\section{Introduction}\label{S:Intro}

The study of discrete integrable systems has been becoming one of the most prominent branches in the theory of integrable systems in the past two decades,
resulting in the establishment of many novel concepts and theories in modern mathematics, see e.g. \cite{HJN16}.
There are several methods of constructing integrable discretisation of nonlinear differential equations.
Among those, a very effective one is to construct the \ac{BT} and the superposition formula of a \ac{PDE},
and consider them as the associated \ac{DDeltaE} and \ac{PDeltaE}, respectively (see \cite{LB80} and \cite{NQC83}).
Compared with integrable \ac{PDE}s, the associated integrable \ac{PDeltaE}s seem to possess a much richer structure,
which reflects in the fact that the discrete equations encode the information of the whole hierarchy of the corresponding continuous equations in an implicit way.

Among the theory of discrete integrable systems, there are three important scalar \ac{PDeltaE}s in three dimensions.
The first one is the discrete \ac{KP} equation (often known as the Hirota equation or the \ac{HM} equation)
\begin{align}\label{dAKP}
 (p_i-p_j)(\rT_{n_h}\tau)(\rT_{n_i}\rT_{n_j}\tau)+(p_j-p_h)(\rT_{n_i}\tau)(\rT_{n_j}\rT_{n_h}\tau)+(p_h-p_i)(\rT_{n_j}\tau)(\rT_{n_h}\rT_{n_i}\tau)=0,
\end{align}
in which the dependent variable $\tau$ is a function of the discrete arguments $n_j$ and the lattice parameters $p_j$ for $j=1,2,\cdots$,
the notation $\rT_{n_j}$ stands for the forward shift operator with respect to the corresponding discrete argument $n_j$, and $p_h$, $p_i$ and $p_j$ are distinct.
This equation was introduced by Hirota \cite{Hir81} as a discrete analogue of the generalised Toda equation,
and here we adopt the form in \eqref{dAKP} which possesses explicit discrete soliton solutions.
Since equation \eqref{dAKP} is associated with the infinite-dimensional algebra $A_\infty$,
here we also refer to it as the discrete AKP equation, in order to distinguish it from the other two discrete \ac{KP}-type equations below.
The second one is the discrete BKP equation (also referred to as the Miwa equation)
\begin{align}\label{dBKP}
 &(p_h-p_i)(p_i-p_j)(p_j-p_h)\tau(\rT_{n_h}\rT_{n_i}\rT_{n_j}\tau)+(p_h+p_i)(p_h+p_j)(p_i-p_j)(\rT_{n_h}\tau)(\rT_{n_i}\rT_{n_j}\tau) \nonumber \\
 &\qquad+(p_i+p_j)(p_i+p_h)(p_j-p_h)(\rT_{n_i}\tau)(\rT_{n_h}\rT_{n_j}\tau)+(p_j+p_h)(p_j+p_i)(p_h-p_i)(\rT_{n_j}\tau)(\rT_{n_h}\rT_{n_i}\tau)=0,
\end{align}
which was introduced by Miwa in \cite{Miw82}.
Although \eqref{dBKP} has an additional term in comparison with \eqref{dAKP},
it is actually a special reduction of the discrete AKP equation (algebraically its associated algebra $B_\infty$ is a sub-algebra of $A_\infty$).
The third model is the discrete CKP equation (also called the hyperdeterminant equation)
\begin{align}\label{dCKP}
(A_1+A_2-A_3-A_4)^2=4B_1B_2,
\end{align}
where the expressions $A_i$ and $B_i$ are given by
\begin{align*}
&A_1=(p_h-p_i)^2(p_i-p_j)^2(p_j-p_h)^2\tau(\rT_{n_h}\rT_{n_i}\rT_{n_j}\tau), \quad A_2=(p_h+p_i)^2(p_h+p_j)^2(p_i-p_j)^2(\rT_{n_h}\tau)(\rT_{n_i}\rT_{n_j}\tau), \\
&A_3=(p_i+p_j)^2(p_i+p_h)^2(p_j-p_h)^2(\rT_{n_i}\tau)(\rT_{n_h}\rT_{n_j}\tau), \quad A_4=(p_j+p_h)^2(p_j+p_i)^2(p_h-p_i)^2(\rT_{n_j}\tau)(\rT_{n_h}\rT_{n_i}\tau), \\
&B_1=(p_h^2-p_i^2)(p_j^2-p_h^2)\left[(p_i+p_j)^2(\rT_{n_i}\tau)(\rT_{n_j}\tau)-(p_i-p_j)^2\tau(\rT_{n_i}\rT_{n_j}\tau)\right], \quad \hbox{and} \quad B_2=\rT_{n_h}B_1,
\end{align*}
respectively, which is also a reduced equation from the discrete AKP equation.
Equation \eqref{dCKP} appeared in Kashaev's paper \cite{Kas96} for the first time from the star-triangle transform in the Ising model,
and it was later identified by Schief \cite{Sch03} that such a model actually describes the superposition formula for the continuous \ac{KP} equation of $C_\infty$-type.
The form in \eqref{dCKP} that contains lattice parameters in the coefficients was given by the authors in \cite{Fu17a} in order to construct its soliton-type solution.

The discrete \ac{KP}-type equations are not only remarkable in their own right in the integrable systems theory,
since these equations, as higher-dimensional models, can reduce to many lower-dimensional integrable discrete systems,
such as the famous discrete \ac{KdV} and \ac{BSQ} systems, see e.g. \cite{DJM3} and \cite{NPCQ92},
but also play crucial roles in other subjects in modern mathematics, especially in geometry.
It was shown by Konopelchenko and Schief that the discrete AKP, BKP and CKP equations are connected to fundamental theorems of plane geometry,
i.e. Menelaus' theorem \cite{KS02a}, Reciprocal quadrangles \cite{KS02b} and Carnot's theorem \cite{Sch03}, respectively.
Besides, Doliwa pointed out that these models also arise in the \ac{MDC} quadrilateral lattice theory in discrete geometry \cite{Dol07,Dol10a,Dol10b}.
Very recently, it was also revealed that the discrete \ac{KP}-type equations are closely related to discrete line complexes \cite{BS15} and circle complexes \cite{BS17}.

In addition to \ac{PDeltaE}s, there are also the so-called generating \ac{PDE}s,
where the terminology `generating' follows from the fact that it generates a whole continuous integrable hierarchy.
From the perspective of integrable systems, the key step of constructing generating \ac{PDE}s is
to take the lattice parameters in the associated \ac{PDeltaE}s as independent variables.
This technique is actually a reflection of introducing Miwa's coordinates, see \cite{KMA99}.
To be more precise, the derivative with respect to a lattice parameter is equivalent to the higher-order derivatives in terms of the continuous flow variables,
resulting in the higher-order symmetries in an integrable hierarchy.
The first example was proposed in \cite{NHJ00} for the \ac{KdV} class in the form of a nonautonomous fourth-order nonlinear equation,
and it is integrable in sense that it possesses soliton, Lax pair, Lagrangian structure, and Painlev\'e reduction, etc.
More importantly, it was shown in \cite{TTX01} that a proper generalisation of this equation
incorporates the hyperbolic Ernst equation for a Weyl neutrino field in general relativity.
In spirit of this, a class of generating {PDE}s for the \ac{BSQ} family, also as integrable nonlinear models, were constructed in \cite{TN05},
and they represent the hyperbolic Ernst equations for a source-free Maxwell field and a Weyl neutrino field, as a generalisation for the previous result of \ac{KdV}.

It turns out that it is very difficult to derive closed-form equations for
a scalar field that would be the analogues of the generating \ac{PDE}s for the \ac{KP}-type hierarchies.
However, as we will show in this paper, it is possible to derive semi-discrete analogues (in form of \ac{DDeltaE}) of
these generating \ac{PDE}s for the KP-type equations,
which simultaneously contain lattice variables and lattice parameters as independent variables.
Such \ac{DDeltaE}s, as nonautonomous semi-discrete equations, are significant in integrable systems theory.
In fact, on the \ac{2D} level, these semi-discrete equations appear in similarity reductions to discrete Painlev\'e equations \cite{NRGO01},
and also play roles of master symmetries of \ac{2D} integrable difference equations, see e.g. \cite{RH07,XNL11}.

We aim to construct the associated nonautonomous \ac{DDeltaE}s for the KP-type equations,
based on the \ac{DL} framework for the KP-type equations given in \cite{Fu17a}.
The \ac{DL} method was proposed by Fokas and Ablowitz (see e.g. \cite{FA81,FA83}) to solve the initial value problems for the \ac{KdV} and \ac{KP} equations,
as a generalisation of the well-known \ac{RHP} \cite{AC91}.
Subsequently, it was developed into a powerful tool to systematically study integrable structures
behind families of discrete and continuous nonlinear equations and their interrelations, see e.g. \cite{NQLC83,NQC83,NCWQ84,NPCQ92,NRGO01,NC90}.
The key idea in the \ac{DL} is to associate a nonlinear equation with a linear integral equation, and by introducing the infinite matrix structure,
it allows us to observe the integrability and solution structures of a nonlinear system simultaneously.
Recently, the link between the linear integral equation and several affine Lie algebras is further established.
With the help of considering reductions on the measure of the linear integral equation,
the \ac{DL} scheme for the discrete AKP, BKP and CKP equations was proposed \cite{Fu17a}.
This makes it possible to further study the associated nonautonomous \ac{DDeltaE}s for these KP-type equations from the general framework.

In this paper, we establish six nonautonomous \ac{DDeltaE}s. They are
\bse
\begin{align}
&\frac{1}{2}(p_i-p_j)^2\rD_{p_i}\rD_{p_j}\tau\cdot\tau=n_in_j\left[(\rT_{n_i}\rT_{n_j}^{-1}\tau)(\rT_{n_i}^{-1}\rT_{n_j}\tau)-\tau^2\right], \label{ddAKP} \\
&(p_i-p_j)\rD_{p_j}(\rT_{n_i}\tau)\cdot\tau=n_j\left[\tau(\rT_{n_i}\tau)-(\rT_{n_j}\tau)(\rT_{n_i}\rT_{n_j}^{-1}\tau)\right], \label{ddmAKP} \\
&(p_i-p_j)\rD_{p_j}[\tau(\rT_{n_i}\rT_{n_j}\tau)]\cdot[(\rT_{n_i}\tau)(\rT_{n_j}\tau)] \nonumber \\
&\qquad=n_j(\rT_{n_j}\tau)^2(\rT_{n_i}\rT_{n_j}\tau)(\rT_{n_i}\rT_{n_j}^{-1}\tau)-(n_j+1)\tau(\rT_{n_i}\tau)^2(\rT_{n_j}^2\tau)
+\tau(\rT_{n_i}\tau)(\rT_{n_j}\tau)(\rT_{n_i}\rT_{n_j}\tau) \label{ddumAKP}
\end{align}
\ese
in the AKP class\footnote{
Here by class we mean a family of equations possess the same solution from the perspective of the \ac{DL}, see also section \ref{S:Integr} for more details.},
\bse
\begin{align}
&2p_ip_j(p_i^2-p_j^2)\rD_{p_i}\rD_{p_j}\tau\cdot\tau \nonumber \\
&\qquad=n_in_j\left[(p_i+q_j)^4(\rT_{n_i}\rT_{n_j}^{-1}\tau)(\rT_{n_i}^{-1}\rT_{n_j}\tau)-(p_i-q_j)^4(\rT_{n_i}\rT_{n_j}\tau)(\rT_{n_i}^{-1}\rT_{n_j}^{-1}\tau)-8p_ip_j(p_i^2+p_j^2)\tau^2\right], \label{ddBKP} \\
&2p_j(p_i^2-p_j^2)\rD_{p_j}(\rT_{n_i}\tau)\cdot\tau=n_j\left[4p_ip_j\tau(\rT_{n_i}\tau)+(p_i-p_j)^2(\rT_{n_j}^{-1}\tau)(\rT_{n_i}\rT_{n_j}\tau)-(p_i+p_j)^2(\rT_{n_j}\tau)(\rT_{n_i}\rT_{n_j}^{-1}\tau)\right] \label{ddmBKP}
\end{align}
\ese
in the BKP class, and
\begin{align}
&p_ip_j(p_i^2-p_j^2)^2\rD_{p_i}\rD_{p_j}\tau\cdot\tau
=n_in_j\left[(p_i-p_j)^2E^{\frac{1}{2}}F^{\frac{1}{2}}+(p_i+p_j)^2G^{\frac{1}{2}}H^{\frac{1}{2}}-8p_ip_j(p_i^2+p_j^2)\tau^2\right] \label{ddCKP}
\end{align}
in the CKP class, where $E$, $F$, $G$ and $H$ are given by
\begin{align*}
&E=(p_i+p_j)^2(\rT_{n_i}^{-1}\tau)(\rT_{n_j}^{-1}\tau)-(p_i-p_j)^2\tau(\rT_{n_i}^{-1}\rT_{n_j}^{-1}\tau), \quad F=(p_i+p_j)^2(\rT_{n_i}\tau)(\rT_{n_j}\tau)-(p_i-p_j)^2\tau(\rT_{n_i}\rT_{n_j}\tau), \\
&G=(p_i-p_j)^2(\rT_{n_i}^{-1}\tau)(\rT_{n_j}\tau)-(p_i+p_j)^2\tau(\rT_{n_i}^{-1}\rT_{n_j}\tau) \quad \hbox{and} \quad
H=(p_i-p_j)^2(\rT_{n_i}\tau)(\rT_{n_j}^{-1}\tau)-(p_i+p_j)^2\tau(\rT_{n_i}\rT_{n_j}^{-1}\tau),
\end{align*}
respectively, where the notation $\rD_{\cdot}$ stands for Hirota's bilinear derivative (see e.g. \cite{Hir04}) with respect to the corresponding arguments $p_i$ and $p_j$,
which is defined by
\begin{align*}
\rD_x f\cdot g=(\partial_x-\partial_{x'})f(x)g(x')|_{x'=x},
\end{align*}
for arbitrary differentiable functions $f(x)$ and $g(x)$.

The tau functions for the above equations have their precise definitions
in terms of infinite matrices and double integral with regard to the spectral variables,
which will be given in the sections of the derivations of these equations.
We note that an autonomous version of \eqref{ddBKP} was given in \cite{Vek19} very recently,
which plays the role of a higher-order semi-discrete BKP equation (i.e. a symmetry) and was referred to as the (2+2)-dimensional Toda lattice.
But here equation \eqref{ddBKP} is a nonautonomous equation having the lattice parameters as the continuous independent variables,
potentially acting as the master symmetry of the discrete BKP equation.

The paper is organised as follows. Section 2 concerns the formal structure of the direct linearisation approach.
In sections \ref{S:AKP}, \ref{S:BKP} and \ref{S:CKP}, we provide derivations of \eqref{ddAKP}, \eqref{ddmAKP} and \eqref{ddumAKP} in the AKP class,
\eqref{ddBKP} and \eqref{ddmBKP}in the BKP class, as well as \eqref{ddCKP} in the CKP class, respectively.
The brief discussion on the integrability of the six \ac{DDeltaE}s is made in section \ref{S:Integr}.

\section{Formal structure of the direct linearisation}\label{S:DL}

In \cite{Fu17a}, the discrete AKP, BKP and CKP equations were studied within a single framework, given by the \ac{DLT},
which provides a dressing-type scheme for obtaining new solutions from given seed solutions for those integrable equations.
In contrast, what we mean by \ac{DL} is a special case where the seed corresponds to a `free' solution (namely when the initial solution is trivial) in the \ac{DLT}.
The latter restriction is useful if we want to derive new equations from some basic assumptions about the initial solutions.

We start with introducing some fundamental infinite matrices and vectors and their properties which are needed in the \ac{DL} framework.
First we need a rank 1 projection matrix
\begin{align}\label{Projection}
 \bO=
 \begin{pmatrix}
 \ddots & & & & \\
 & 0 & & \\
 & & \boxed{1} & & \\
 & & & 0 & \\
 & & & & \ddots
 \end{pmatrix},
\end{align}
where the `box' denotes the location of the central element, namely the $(0,0)$-entry of the matrix.
This infinite matrix has the property
\begin{align*}
 (\bO \bU)_{i,j}=\delta_{i,0}U_{0,j} \quad \hbox{and} \quad (\bU \bO)_{i,j}=U_{i,0}\delta_{0,j}, \quad \hbox{with} \quad
 \delta_{i,j}=
 \left\{
 \begin{array}{ll}
 1, & i=j, \\
 0, & i\neq j,
 \end{array}
 \right. \quad
 \forall i,j\in\mathbb{Z},
\end{align*}
for an arbitrary infinite matrix
\begin{align*}
 \bU=
 \begin{pmatrix}
  & \vdots & \vdots & \vdots & \\
 \cdots & U_{-1,-1} & U_{-1,0} & U_{-1,1} & \cdots\\
 \cdots & U_{0,-1} & \boxed{U_{0,0}} & U_{0,1} & \cdots\\
 \cdots & U_{1,-1} & U_{1,0} & U_{1,1} & \cdots\\
 & \vdots & \vdots & \vdots &
 \end{pmatrix},
\end{align*}
where the operation $(\cdot)_{i,j}$ denotes taking the $(i,j)$-entry.

Next, we introduce two infinite matrices
\begin{align}\label{Index}
 \bLd=
 \begin{pmatrix}
 \ddots & \ddots & & & \\
 & 0 & 1 & \\
 & & \boxed{0} & 1 & \\
 & & & 0 & \ddots \\
 & & & & \ddots
 \end{pmatrix}
 \quad \hbox{and} \quad
 \tbLd=
 \begin{pmatrix}
 \ddots & & & & \\
 \ddots & 0 & & & \\
 & 1 & \boxed{0} & & \\
 & & 1 & 0 & \\
 & & & \ddots & \ddots
 \end{pmatrix}.
\end{align}
The matrices $\bLd$ and $\tbLd$ are the transpose of each other, and have properties
\begin{align*}
 (\bLd\,\bU)_{i,j}=U_{i+1,j} \quad \hbox{and} \quad (\bU\,\tbLd)_{i,j}=U_{i,j+1},
\end{align*}
namely the multiplications by $\bLd$ and $\tbLd$ raise the row index and the column index of $\bU$, respectively.
For this reason, we refer to them as index-raising matrices.
We also introduce the infinite-dimensional vectors
\begin{align}\label{bc}
 \bc_k=(\cdots,k^{-1},1,k,\cdots)^\rT \quad \hbox{and} \quad
 \tbc_{k'}=(\cdots,k'^{-1},1,k',\cdots),
\end{align}
which obey the following identities:
\begin{align*}
 \bLd\,\bc_k=k\,\bc_k, \quad \tbc_{k'}\tbLd=k'\tbc_{k'}.
\end{align*}
In the sections below, we also need the notions of trace and determinant for an infinite matrix.
We give the formal definition of a trace of an arbitrary infinite matrix $\bU$ as follows:
\begin{align}\label{Trace}
 \tr\bU=\sum_{i\in\mathbb{Z}}U_{i,i}.
\end{align}
Notice that this is a formal definition, since the infinite summation with respect to $i$ over the integer ring may lead to divergence.
In order to avoid this issue, in this paper we only deal with the trace of a matrix involving the projection matrix $\bO$,
in which case the trace is always convergent. For example, we have
\begin{align*}
\tr(\bO\bU)=\tr(\bU\bO)=U_{0,0}=(\bU)_{0,0}.
\end{align*}
In the convergent case, the trace also satisfies $\tr(\bU\bV)=\tr(\bV\bU)$ for arbitrary infinite matrices $\bU$ and $\bV$.
The determinant of an infinite matrix is defined through
\begin{align}\label{Det}
 \ln(\det\bU)=\tr(\ln\bU).
\end{align}
Again, this is a formal definition which could result in divergence problem.
But if we restrict ourselves to the infinite matrix $1+\bU$,
where $1$ is the identity infinite matrix and $\bU$ is an infinite matrix involving the projection matrix $\bO$,
the determinant is well-defined, since the right hand side of \eqref{Det}, namely
\begin{align*}
 \tr\left[\ln(1+\bU)\right]=\tr\left[\sum_{i=1}^\infty\frac{(-1)^{i-1}}{i}\bU^i\right]=\sum_{i=1}^\infty\frac{(-1)^{i-1}}{i}\tr\left(\bU^i\right),
\end{align*}
has terms of convergent traces. We have the well-known Weinstein--Aronszajn formulas for the determinant. For instance, the following identities hold:
\begin{align*}
 \det(1+\bU\bO\bV)=1+(\bV\bU)_{0,0}, \quad
 \det(1+\bU(\bO\bLd-\tbLd\bO)\bV)
 =\det
 \begin{pmatrix}
  1+(\bLd\bV\bU)_{0,0} & -(\bLd\bV\tbLd)_{0,0} \\
  (\bV\bU)_{0,0} & 1-(\bV\bU\tbLd)_{0,0}
 \end{pmatrix}.
\end{align*}
which are the cases of rank 1 and rank 2.

We provide the formal structure of the \ac{DL} framework. The starting point is a linear integral equation
\begin{align}\label{Integral}
\bu_k+\iint_D\rd\zeta(l,l')\rho_k\Oa_{k,l'}\sigma_{l'}\bu_l=\rho_k\bc_k,
\end{align}
in which the wave function $\bu_k$ is an infinite-dimensional column vector having its $i$-th component $u_k^{(i)}$ (for $i\in\mathbb{Z}$)
being a function of the lattice variables $n_j$ and the lattice parameters $p_j$ for $j=1,2,\cdots$ as well as the spectral variable $k$,
$\Oa_{k,l'}$ is the kernel of the linear integral equation, depending on the spectral variables $k$ and $l'$,
$\rd\zeta$ and $D$ are the measure and the domain for integration,
and $\rho_k$ and $\sigma_{l'}$ are the so-called plane wave factors depending on the discrete variables $n_j$ and lattice parameters $p_j$
as well as their respective spectral variables $k$ and $l'$.
Furthermore, to derive equations in given choice of independent variables (which could be either the discrete variables $n_j$ or the continuous variables $p_j$)
we assume that the measure is independent of these chosen variables.

We present the infinite matrix representation of \eqref{Integral}. We first introduce the infinite matrix $\bOa$ defined through
\begin{align}\label{Omega}
\Oa_{k,k'}=\tbc_{k'}\bOa\,\bc_k.
\end{align}
Following from the properties of the index-raising matrices and the projection matrix listed above,
we can observe that $\bOa$ is actually an infinite matrix composed of $\bLd$, $\tbLd$ and $\bO$, which relies on the precise expression of $\Oa_{k,k'}$,
namely it is an infinite matrix representation of the Cauchy kernel of the linear integral equation \eqref{Integral}.
If we replace $\Oa_{k,l'}$ with the help of \eqref{Omega}, the linear integral equation is reformulated as
\begin{align}\label{uk}
\bu_k=(1-\bU\bOa)\bc_k\rho_k,
\end{align}
where the infinite matrix $\bU$ is given by
\begin{align}\label{Potential}
\bU\doteq\iint_D\rd\zeta(k,k')\bu_k\tbc_{k'}\sigma_{k'}.
\end{align}
Next, we consider the infinite matrix representation of the plane wave factors and introduce an infinite matrix
\begin{align}\label{C}
\bC\doteq\iint_D\rd\zeta(k,k')\rho_k\bc_k\tbc_{k'}\sigma_{k'}.
\end{align}
The key characteristic of $\bC$ is the product of the two plane wave factors, i.e. $\rho_k\sigma_{k'}$,
which we normally refer to as the so-called effective plane wave factor.
By acting the operation $\iint_D\rd\zeta(k,k')\cdot\tbc_{k'}\sigma_{k'}$ on equation \eqref{C}, we obtain
\begin{align}\label{U}
\bU=(1-\bU\bOa)\bC, \quad \hbox{or equivalently} \quad \bU=\bC(1+\bOa\bC)^{-1}.
\end{align}

The idea of the \ac{DL} approach is to associate a nonlinear equation with a linear integral equation in the form of \eqref{Integral}.
Once the plane wave factors, the Cauchy kernel and the measure are given, the corresponding class of nonlinear integrable systems is fully determined.
To be more precise, for a certain class of nonlinear integrable equations, the infinite matrix $\bC$ describes the linear dispersion,
the infinite matrix $\bOa$ together with the measure governs the nonlinear structure of the corresponding integrable models,
and $\bu_k$ and $\bU$ are corresponding to the wave function in the Lax pair and the nonlinear potential, respectively.

\section{Derivation of \eqref{ddAKP}, \eqref{ddmAKP} and \eqref{ddumAKP}}\label{S:AKP}

In the AKP class, the discrete plane wave factors $\rho_k$ and $\sigma_{k'}$ are given by
\begin{align}\label{A:PWF}
\rho_{k}=\prod_{i=1}^\infty(p_i+k)^{n_i} \quad \hbox{and} \quad \sigma_{k'}=\prod_{i=1}^\infty(p_i-k')^{-n_i},
\end{align}
respectively. Substituting these into the infinite matrix $\bC$ defined by \eqref{C}
and considering the evolutions with regard to the discrete variables and lattice parameters,
we obtain dynamical evolutions as follows:
\bse\label{A:CDyn}
\begin{align}
&(\rT_{n_j}\bC)(p_j-\tbLd)=(p_j+\bLd)\bC, \label{A:CDyna} \\
&\partial_{p_j}\bC=n_j\left(\frac{1}{p_j+\bLd}\bC-\bC\frac{1}{p_j-\tbLd}\right), \label{A:CDynb} \\
&(\rT_a^{-1}\rT_b\bC)\frac{b-\tbLd}{a-\tbLd}=\frac{b+\bLd}{a+\bLd}\bC. \label{A:CDync}
\end{align}
\ese
Below we provide the derivation of \eqref{A:CDynb}.
By differentiating $\rho_k\sigma_{k'}$ with respect to $p_j$, we have
\begin{align*}
\partial_{p_j}\rho_k\sigma_{k'}=\partial_{p_j}\prod_{i=1}^\infty\left(\frac{p_i+k}{p_i-k'}\right)^{n_i}
=n_j\left(\frac{1}{p_j+k}-\frac{1}{p_j-k'}\right)\prod_{i=1}^\infty\left(\frac{p_i+k}{p_i-k'}\right)^{n_i}=n_j\left(\frac{1}{p_j+k}-\frac{1}{p_j-k'}\right)\rho_k\sigma_{k'},
\end{align*}
and thus, the same operation on $\bC$ gives rise to
\begin{align*}
\partial_{p_j}\bC={}&\iint_{D}\rd\zeta(k,k')\bc_k\partial_{p_j}(\rho_k\sigma_{k'})\tbc_{k'}
=\iint_{D}\rd\zeta(k,k')\bc_k\partial_{p_j}(\rho_k\sigma_{k'})\tbc_{k'} \\
={}&n_j\left(\iint_{D}\rd\zeta(k,k')\frac{1}{p_j+k}\bc_k\rho_k\sigma_{k'}\tbc_{k'}-\iint_{D}\rd\zeta(k,k')\bc_k\rho_k\sigma_{k'}\tbc_{k'}\frac{1}{p_j-k'}\right)
=n_j\left(\frac{1}{p_j+\bLd}\bC-\bC\frac{1}{p_j-\tbLd}\right),
\end{align*}
where in the last step we have made use of the property of the operation of the index-raising matrices on $\bc_k$ and $\tbc_{k'}$ given in section \ref{S:DL}.
The other two equations are derived similarly.

The Cauchy kernel in the linear integral equation for the discrete AKP equation takes the form of
\begin{align}\label{A:Kernel}
\Oa_{k,k'}=\frac{1}{k+k'},
\end{align}
and in this case we have no further requirement for the measure $\rd\zeta(k,k')$ and the domain $D$, namely they are arbitrary.
According to \eqref{Omega}, we can observe that in this case $\bOa=-\sum_{i=0}^\infty(-\tbLd)^{-i-1}\bO\bLd^i$;
in other words, it satisfies
\begin{align}\label{A:Omega}
\bOa\bLd+\tbLd\bOa=\bO.
\end{align}
Equation \eqref{A:Omega} can also be written in other forms. Here we reformulate it in the following forms whose left hand sides are compatible with \eqref{A:CDyn}:
\bse\label{A:OmegaDyn}
\begin{align}
&\bOa(p_j+\bLd)+(p_j-\tbLd)\bOa=\bO, \label{A:OmegaDyna}\\
&\bOa\frac{1}{p_j+\bLd}-\frac{1}{p_j-\tbLd}\bOa=\frac{1}{-p_j+\tbLd}\bO\frac{1}{p_j+\bLd}, \label{A:OmegaDynb}\\
&\bOa\frac{b+\bLd}{a+\bLd}-\frac{b-\tbLd}{a-\tbLd}\bOa=-(a-b)\frac{1}{-a+\tbLd}\bO\frac{1}{a+\bLd}. \label{A:OmegaDync}
\end{align}
\ese

Equations \eqref{A:OmegaDyn} together with \eqref{A:CDyn} will provide the dynamical evolutions of the infinite matrix $\bU$ as follows:
\bse\label{A:UDyn}
\begin{align}
&(\rT_{n_j}\bU)(p_j-\tbLd)=(p_j+\bLd)\bU-(\rT_{n_j}\bU)\bO\bU, \label{A:UDyna} \\
&\partial_{p_j}\bU=n_j\left(\frac{1}{p_j+\bLd}\bU-\bU\frac{1}{p_j-\tbLd}
-\bU\frac{1}{-p_j+\tbLd}\bO\frac{1}{p_j+\bLd}\bU\right), \label{A:UDynb} \\
&\left(\rT_a\rT_b^{-1}\bU\right)\frac{b-\tbLd}{a-\tbLd}=\frac{a+\bLd}{b+\bLd}\bU
+(a-b)(\rT_a^{-1}\rT_b\bU)\frac{1}{-a+\tbLd}\bO\frac{1}{a+\bLd}\bU. \label{A:UDync}
\end{align}
\ese
Again we only give the derivation of \eqref{A:UDynb} below and skip that of the other two, as the procedure is similar.
Notice that the infinite matrix $\bU$ obeys \eqref{U} in the formal structure of the \ac{DL}.
Calculating the derivative of $\bU$ with respect to $p_j$, we obtain
\begin{align*}
\partial_{p_j}\bU=-(\partial_{p_j}\bU)\bOa\bC+(1-\bU\bOa)(\partial_{p_j}\bC),
\end{align*}
which can equivalently be rewritten as
\begin{align*}
(\partial_{p_j}\bU)(1+\bOa\bC)={}&n_j(1-\bU\bOa)\left(\frac{1}{p_j+\bLd}\bC-\bC\frac{1}{p_j-\tbLd}\right) \\
={}&n_j\left[\frac{1}{p_j+\bLd}\bC-\bU\frac{1}{p_j-\tbLd}-\bU\bOa\frac{1}{p_j+\bLd}\bC\right] \\
={}&n_j\left[\frac{1}{p_j+\bLd}\bC-\bU\frac{1}{p_j-\tbLd}(1+\bOa\bC)-\bU\frac{1}{-p_j+\tbLd}\bO\frac{1}{p_j+\bLd}\bC\right],
\end{align*}
where in the first and third equalities we have used \eqref{A:CDynb} and \eqref{A:OmegaDynb}, respectively.
Multiplying the above equation by $(1+\bOa\bC)^{-1}$ immediately gives rise to \eqref{A:UDynb}.

Next, we introduce the tau function in this class, which is defined as
\begin{align}\label{A:tau}
\tau\doteq\det(1+\bOa\bC),
\end{align}
where $1$ denotes the infinite unit matrix.
As we have mentioned in section \ref{S:DL}, the determinant of an infinite matrix should be understood as the formal expansion of $\exp\{\tr[\ln(1+\bOa\bC)]\}$.
Since $\bOa=-\sum_{i=0}^\infty(-\tbLd)^{-i-1}\bO\bLd^i$ involves $\bO$ in every term, the trace action is always convergent.
For convenience, we introduce quantities
\begin{align*}
u\doteq (U)_{0,0}, \quad V_a\doteq 1-\left(\bU\frac{1}{a+\tbLd}\right)_{0,0}, \quad W_a\doteq 1-\left(\frac{1}{a+\bLd}\bU\right)_{0,0} \quad \hbox{and} \quad
S_{a,b}\doteq \left(\frac{1}{a+\bLd}\bU\frac{1}{b+\tbLd}\right)_{0,0}.
\end{align*}
The tau function satisfies the dynamical evolutions with respect to the discrete variables and the lattices parameters as follows:
\bse\label{A:tauDyn}
\begin{align}
&\frac{\rT_{n_j}\tau}{\tau}=1-\left(\bU\frac{1}{-p_j+\tbLd}\right)_{0,0}=V_{-p_j}, \quad
\frac{\rT_{n_j}^{-1}\tau}{\tau}=1-\left(\frac{1}{p_j+\bLd}\bU\right)_{0,0}=W_{p_j}, \label{A:tauDyna} \\
&\partial_{p_j}\ln\tau=n_jS_{p_j,-p_j}, \label{A:tauDynb} \\
&\frac{\rT_a^{-1}\rT_b\tau}{\tau}=1-(a-b)\left(\frac{1}{a+\bLd}\bU\frac{1}{-b+\tbLd}\right)_{0,0}=1-(a-b)S_{a,-b}, \label{A:tauDync} \\
&(p_i-p_j)\frac{\tau(\rT_{n_i}\rT_{n_j}\tau)}{(\rT_{n_i}\tau)(\rT_{n_j}\tau)}=p_i-p_j+\rT_{n_j}u-\rT_{n_i}u. \label{A:tauDynd}
\end{align}
\ese
These equations are proven through direct computation in terms of infinite matrices.
For instance, calculating the derivative of $\ln\tau$ with respect to $p_j$ yields
\begin{align*}
\partial_{p_j}\ln\tau={}&\partial_{p_j}\ln[\det(1+\bOa\bC)]=\partial_{p_j}\tr[\ln(1+\bOa\bC)]=\tr[\partial_{p_j}\ln(1+\bOa\bC)]=\tr[(1+\bOa\bC)^{-1}\bOa(\partial_{p_j}\bC)] \\
={}&n_j\tr\left[(1+\bOa\bC)^{-1}\bOa\left(\frac{1}{p_j+\bLd}\bC-\bC\frac{1}{p_j-\tbLd}\right)\right]
=n_j\tr\left[\bC(1+\bOa\bC)^{-1}\left(\bOa\frac{1}{p_j+\bLd}-\frac{1}{p_j-\tbLd}\bOa\right)\right].
\end{align*}
where in the last step the property of the cyclic permutation of the trace operation is used.
With the help of \eqref{U} and \eqref{A:OmegaDynb}, this equation is reformulated as
\begin{align*}
\partial_{p_j}\ln\tau=n_j\tr\left(\bU\frac{1}{-p_j+\tbLd}\bO\frac{1}{p_j+\bLd}\right)
=n_j\tr\left(\bO\frac{1}{p_j+\bLd}\bU\frac{1}{-p_j+\tbLd}\right)=n_j\left(\frac{1}{p_j+\bLd}\bU\frac{1}{-p_j+\tbLd}\right)_{0,0},
\end{align*}
namely $\partial_{p_j}\ln\tau=n_jS_{p_j,-p_j}$.
To prove Equations \eqref{A:tauDyna} and \eqref{A:tauDync},
one needs to use \eqref{A:CDyna} and \eqref{A:OmegaDyna} and \eqref{A:CDync} and \eqref{A:OmegaDync}, respectively.
Since the idea of the proofs is similar to that of \eqref{A:tauDyna}, we skip them here.
While equation \eqref{A:tauDynd} is proven based on \eqref{A:tauDyna} and \eqref{A:UDyn},
and it is a widely known relation which describes bilinear transformation for the discrete \ac{KP} equation (see e.g. \cite{Fu17a}).

Equations listed in \eqref{A:UDyn} are the key formulas in the \ac{DL} scheme to construct closed-form integrable equations,
and they together with \eqref{A:tauDyn} can produce the bilinear equations in the AKP class.
In \cite{Fu17a}, the well-known \ac{HM} equation, i.e. equation \eqref{dAKP}, is derived from \eqref{A:UDyna} and \eqref{A:tauDyna}.
Here we start from the dynamical relations in terms of $p_j$ to construct the (2+2)-dimensional nonautonomous \ac{DDeltaE} in this class.
Considering $(a+\bLd)^{-1}\eqref{A:UDynb}(b+\tbLd)^{-1}$ and taking the $(0,0)$-entry, we have the following equation for the quantity $S_{a,b}$:
\begin{align*}
\partial_{p_j}S_{a,b}=n_j\left[\left(\frac{1}{p_j-a}-\frac{1}{p_j+b}\right)S_{a,b}-\frac{1}{p_j-a}S_{p_j,b}+\frac{1}{p_j+b}S_{a,-p_j}-S_{a,-p_j}S_{p_j,b}\right].
\end{align*}
By setting $a=p_i$ and $b=-p_i$, we reach to
\begin{align*}
\partial_{p_j}S_{p_i,-p_i}=\frac{n_j}{(p_i-p_j)^2}\left[\left(1-(p_j-p_i)S_{p_j,-p_i}\right)\left(1-(p_i-p_j)S_{p_i,-p_j}\right)-1\right],
\end{align*}
which only involves the $S$-variable. If we replace all the $S$-variables in this equation by the tau function via \eqref{A:tauDyn}
and notice the identity
\begin{align*}
\partial_{p_i}\partial_{p_j}\ln\tau=\frac{\rD_{p_i}\rD_{p_j}\tau\cdot\tau}{2\tau^2}
\end{align*}
for Hirota's bilinear operator $\rD_\cdot$, the bilinear equation \eqref{ddAKP} arises.
This equation takes the form of a nonautonomous version of the \ac{2D} Toda equation,
if we think of the discrete shift operations $\rT_{n_i}\rT_{n_j}^{-1}$ and $\rT_{n_i}^{-1}\rT_{n_j}$ as the forward and backward shifts along skew direction on the lattice.
The difference is that here the lattice parameters $p_i$ and $p_j$ act the independent variables;
while in the \ac{2DTL}, the bilinear derivatives are with respect to the continuous flow variables $x_1$ and $x_{-1}$ (cf. \cite{Fu18a}),
which leads to an autonomous equation.

From the scheme, we can also construct two more nonautonomous semi-discrete equations of $\tau$.
Taking $[\eqref{A:UDynb}(a+\tbLd)^{-1}]_{0,0}$, one obtains
\begin{align*}
\partial_{p_j}V_a=-n_j\left[V_{-p_j}S_{p_j,a}+\frac{1}{a+p_j}(V_a-V_{-p_j})\right],
\end{align*}
and its $a=-p_i$ case gives rise to
\begin{align*}
\partial_{p_j}V_{-p_i}=\frac{n_j}{p_i-p_j}\left[V_{-p_i}-V_{-p_j}\left(1-(p_j-p_i)S_{p_j,-p_i}\right)\right].
\end{align*}
Substituting the $V$- and $S$-variables with the tau function with the help of \eqref{A:tauDyn}, we are able to construct the bilinear equation \eqref{ddmAKP}.

Moreover, if we consider the equation for $\partial_{p_j}u$ by evaluating the $(0,0)$-entry of \eqref{A:UDynb}, the following equation shows up:
\begin{align*}
\partial_{p_j}u=n_j(1-V_{-p_j}W_{p_j})=n_j\left[1-\frac{(\rT_{n_j}\tau)(\rT_{n_j}^{-1}\tau)}{\tau^2}\right],
\end{align*}
where formula \eqref{A:tauDyn} is used for the second equality,
and therefore, we obtain from \eqref{A:tauDynd} that
\begin{align*}
\partial_{p_j}\left[(p_i-p_j)\frac{\tau(\rT_{n_i}\rT_{n_j}\tau)}{(\rT_{n_i}\tau)(\rT_{n_j}\tau)}\right]=\partial_{p_j}(p_i-p_j+\rT_{n_j}u-\rT_{n_i}u)
=-1+\rT_{n_j}(\partial_{p_j}u)-\rT_{n_i}(\partial_{p_j}u).
\end{align*}
Finally, by substituting $u$ with $\tau$, a closed-form quartic equation of $\tau$ is constructed, which takes the form of equation \eqref{ddumAKP}.

\section{Derivation of \eqref{ddBKP} and \eqref{ddmBKP}}\label{S:BKP}

We select the following plane wave factors for the BKP class:
\begin{align}\label{B:PWF}
\rho_{k}=\prod_{i=1}^\infty\left(\frac{p_i+k}{p_i-k}\right)^{n_i}, \quad \sigma_{k'}=\rho_{k'}=\prod_{i=1}^\infty\left(\frac{p_i+k'}{p_i-k'}\right)^{n_i},
\end{align}
which should be understood as discrete odd flows, compared with \eqref{A:PWF} in the AKP case.
In this case, it is observed that the infinite matrix $\bC$ defined in \eqref{C} evolves with regard to the discrete variables and the lattice parameters in the following way:
\bse\label{B:CDyn}
\begin{align}
&(\rT_{n_j}\bC)\frac{p_j-\tbLd}{p_j+\tbLd}=\frac{p_j+\bLd}{p_j-\bLd}\bC, \label{B:CDyna} \\
&\partial_{p_j}\bC=n_j\left[\left(\frac{1}{p_j+\bLd}-\frac{1}{p_j-\bLd}\right)\bC+\bC\left(\frac{1}{p_j+\tbLd}-\frac{1}{p_j-\tbLd}\right)\right]. \label{B:CDynb}
\end{align}
\ese
The derivation of these formulas is straightforward, as they follow from the formal definition of $\bC$, namely \eqref{C}.

The Cauchy kernel in the BKP case takes the form of
\begin{align}\label{B:Kernel}
\Oa_{k,k'}=\frac{1}{2}\frac{k-k'}{k+k'},
\end{align}
which implies that the corresponding $\bOa$ satisfies
\begin{align}\label{B:Omega}
\bOa\bLd+\tbLd\bOa=\frac{1}{2}(\bOa\bLd-\tbLd\bO).
\end{align}
For the purpose of looking for compatible relations of $\bOa$ with equations listed in \eqref{B:CDyn}, we reformulate \eqref{B:Omega} as
\bse\label{B:OmegaDyn}
\begin{align}\label{B:OmegaDyna}
\bOa\frac{p_j+\bLd}{p_j-\bLd}-\frac{p_j-\tbLd}{p_j+\tbLd}\bOa
=p_j\frac{1}{p_j+\tbLd}(\bO\bLd-\tbLd\bO)\frac{1}{p_j-\bLd}
\end{align}
and
\begin{align}\label{B:OmegaDynb}
&\bOa\left(\frac{1}{p_j+\bLd}-\frac{1}{p_j-\bLd}\right)+\left(\frac{1}{p_j+\tbLd}-\frac{1}{p_j-\tbLd}\right)\bOa \nonumber \\
&\qquad=-\frac{1}{2}\frac{1}{p_j-\tbLd}(\bO\bLd-\tbLd\bO)\frac{1}{p_j+\bLd}
-\frac{1}{2}\frac{1}{p_j+\tbLd}(\bO\bLd-\tbLd\bO)\frac{1}{p_j-\bLd},
\end{align}
\ese
respectively.

Equations \eqref{B:CDyn} and \eqref{B:OmegaDyn} are the fundamental relations to construct the dynamical evolutions of the infinite matrix $\bU$.
By considering $\rT_{n_j}U$ and $\partial_{p_j}\bU$ in \eqref{U}, some straightforward calculation yields the following equations:
\bse\label{B:UDyn}
\begin{align}
&(\rT_{n_j}\bU)\frac{p_j-\tbLd}{p_j+\tbLd}=\frac{p_j+\bLd}{p_j-\bLd}\bU
 -p_j(\rT_{n_j}\bU)\frac{1}{p_j+\tbLd}(\bO\bLd-\tbLd\bO)\frac{1}{p_j-\bLd}\bU, \label{B:UDyna} \\
&\partial_{p_j}\bU=n_j\left[\left(\frac{1}{p_j+\bLd}-\frac{1}{p_j-\bLd}\right)\bU+\bU\left(\frac{1}{p_j+\tbLd}-\frac{1}{p_j-\tbLd}\right)\right. \nonumber \\
&\qquad\qquad\qquad+\left.\frac{1}{2}\bU\frac{1}{p_j-\tbLd}(\bO\bLd-\tbLd\bO)\frac{1}{p_j+\bLd}\bU+\frac{1}{2}\bU\frac{1}{p_j+\tbLd}(\bO\bLd-\tbLd\bO)\frac{1}{p_j-\bLd}\bU\right], \label{B:UDynb}
\end{align}
\ese
with the help of \eqref{B:CDyn} and \eqref{B:OmegaDyn}.

Besides the different kernel and plane wave factors, we also have to impose certain restrictions on the integration measure and the integration domain for the BKP equation.
We require that the integration domain $D$ is symmetric in terms of the spectral variables $k$ and $k'$, and then the measure $\rd\zeta(k,k')$ is antisymmetric, i.e.
\begin{align}\label{B:Reduction}
\rd\zeta(k',k)=-\rd\zeta(k,k').
\end{align}
These will together result in
\begin{align*}
\tbC=\iint_D\rd\zeta(k,k')\rho_k\bc_{k'}\tbc_k\rho_{k'}=-\iint_D\rd\zeta(k',k)\rho_{k'}\bc_{k'}\tbc_k\rho_k=-\bC;
\end{align*}
in other words, we have an antisymmetric infinite matrix $\bC$ in the discrete BKP.
Notice that the kernel given in \eqref{B:Kernel} is also antisymmetric, or equivalently $\tbOa=-\bOa$.
We can deduce from \eqref{U} that in this case the infinite matrix $\bU$ obeys the antisymmetry property
\begin{align}\label{B:USym}
\tbU=-\bU.
\end{align}

Next, we present the dynamical evolutions of the tau function. For convenience we introduce quantities
\begin{align*}
V_a\doteq 1-\left(\bU\frac{a}{a-\tbLd}\right)_{0,0}, \quad W_a\doteq 1+\left(\frac{a}{a-\bLd}\bU\right)_{0,0} \quad \hbox{and} \quad
S_{a,b}\doteq \left(\frac{a}{a-\bLd}\bU\frac{b}{b-\tbLd}\right)_{0,0}.
\end{align*}
Due to the antisymmetry property of the infinite matrix $\bU$, it is obvious to see that
\begin{align*}
V_a=W_a \quad \hbox{and} \quad S_{a,b}=-S_{b,a}.
\end{align*}
We define the tau function in this class by
\begin{align}\label{B:tau}
\tau^2=\det(1+\bOa\bC),
\end{align}
since the antisymmetry of $\bO$ and $\bC$ will eventually make the above determinant a perfect square, namely the tau function itself is corresponding to a Pfaffian.
\bse\label{B:tauDyn}
We now consider the evolution of the tau function with respect to $n_j$, and this gives us
\begin{align*}
 \rT_{n_j}\tau^2={}&\det\left[1+\bOa(\rT_{n_j}\bC)\right]
 =\det\left[1+\bOa\bC+p_j\frac{1}{p_j-\tbLd}(\bO\bLd-\tbLd\bO)\frac{1}{p_j-\bLd}\bC\right] \nonumber \\
 ={}&\tau^2\det
 \left[
 \begin{array}{cc}
 1+p(\frac{\bLd}{p-\bLd}\bU\frac{1}{p-\tbLd})_{0,0} & -p(\frac{\bLd}{p-\bLd}\bU\frac{\tbLd}{p-\tbLd})_{0,0}\\
 p(\frac{1}{p-\bLd}\bU\frac{1}{p-\tbLd})_{0,0} & 1-p(\frac{1}{p-\bLd}\bU\frac{\tbLd}{p-\tbLd})_{0,0}
 \end{array}
 \right]
 =\tau^2V_{p_j}^2,
\end{align*}
where the rank 2 Weinstein--Aronszajn formula and equation \eqref{B:USym} are used in the third and fourth equalities, respectively.
Similarly, acting the backward shift on $\tau$ gives us $\rT_{n_j}^{-1}\tau^2/\tau=V_{-p_j}^2$.
In order to simplify this formula, we evaluate $[\eqref{B:UDyna}]_{0,0}$, which gives us $V_{p_j}(\rT_{n_j} V_{-p_j})=1$.
And thus, without loss of generality, we have
\begin{align}\label{B:tauDyna}
 \frac{\rT_{n_j}\tau}{\tau}=V_{p_j} \quad \hbox{and} \quad \frac{\rT_{n_j}^{-1}\tau}{\tau}=V_{-p_j}.
\end{align}
Next, we calculate $[\frac{p_i}{p_i-\bLd}\eqref{B:UDyna}]_{0,0}$. The following equation is obtained:
\begin{align*}
 1+2V_{p_j}(\rT_{n_j}S_{-p_i,-p_j})=\frac{p_i-p_j}{p_i+p_j}(V_{p_j}-V_{-p_i})+V_{p_j}(\rT_{n_j}V_{-p_i}).
\end{align*}
This equation provides a way to express the $S$-variable by the $V$-variables,
and consequently, the $S$-variable can be expressed by the tau function by making use of \eqref{B:tauDyn}, and the formula is
\begin{align}\label{B:tauDynb}
S_{-p_i,-p_j}=\frac{1}{2}\left[\frac{\rT_{n_i}^{-1}\tau-\rT_{n_j}^{-1}\tau}{\tau}+\frac{p_i-p_j}{p_i+p_j}\left(1-\frac{\rT_{n_i}^{-1}\rT_{n_j}^{-1}\tau}{\tau}\right)\right].
\end{align}
The expressions $S_{p_i,-p_j}$, $S_{-p_i,p_j}$ and $S_{p_i,p_j}$ in terms of the tau function can also be derived from the above equation,
with the help of $(p_i,\rT_{n_i})\leftrightarrow(-p_i,\rT_{n_i}^{-1})$.
Therefore, we have obtained the dynamical relations of $\tau$ in terms of $S_{a,b}$ for arbitrary $a+b\neq0$.
Furthermore, following the same idea of deriving \eqref{A:tauDynb}, we also have
\begin{align}\label{B:tauDync}
2p_j\partial_{p_j}\ln\tau=n_j(V_{p_j}-V_{-p_j}-2S_{p_j,-p_j})
\end{align}
for BKP. Equations listed in \eqref{B:tauDyn} establish the relations between the dynamics of the tau function and the $V$- and $S$-variables.
\ese

To construct equation \eqref{ddBKP} in the \ac{DL} framework, we compute $[\frac{a}{a-\bLd}\eqref{B:UDynb}]_{0,0}$ and $[\frac{a}{a-\bLd}\eqref{B:UDynb}\frac{b}{b-\tbLd}]_{0,0}$,
which gives rise to
\bse
\begin{align}\label{B:VDyn}
p_j\partial_{p_j}V_a=n_j\left[V_a\left(\frac{2ap_j}{a^2-p_j^2}-\frac{1}{2}V_{p_j}+\frac{1}{2}V_{-p_j}\right)+V_{p_j}\left(\frac{1}{2}\frac{p_j+a}{p_j-a}+S_{a,-p_j}\right)-V_{-p_j}\left(\frac{1}{2}\frac{p_j-a}{p_j+a}+S_{a,p_j}\right)\right],
\end{align}
and
\begin{align}\label{B:SDyn}
p_j\partial_{p_j}S_{a,b}={}&n_j\left[\frac{1}{2}S_{p_j,b}\left(\frac{p_j+a}{p_j-a}-W_a\right)+\frac{1}{2}S_{-p_j,b}\left(\frac{a-p_j}{a+p_j}+W_a\right)\right. \nonumber \\
&\qquad+\frac{1}{2}S_{a,p_j}\left(\frac{p_j+b}{p_j-b}-V_b\right)+\frac{1}{2}S_{a,-p_j}\left(\frac{b-p_j}{b+p_j}+V_b\right) \nonumber \\
&\qquad\qquad\left.+S_{a,b}\left(\frac{2ap_j}{a^2-p_j^2}+\frac{2bp_j}{b^2-p_j^2}\right)-S_{a,p_j}S_{-p_j,b}+S_{a,-p_j}S_{p_j,b}\right],
\end{align}
\ese
respectively.
These two equations allow us to derive the closed-form equations in terms of the tau function.
Notice that the logarithm derivative of $\tau$ with respect to $p_j$ satisfies \eqref{B:tauDync},
and hence we have
\begin{align*}
p_ip_j\partial_{p_i}\partial_{p_i}\ln\tau=p_j\partial_{p_j}(p_i\partial_{p_i}\ln\tau)
=p_j\partial_{p_j}[n_i(V_{p_i}-V_{-p_i}-2S_{p_i,p_i})]=n_i\left[p_j(\partial_{p_j}V_{p_i})-p_j(\partial_{p_j}V_{-p_i})-2p_j(\partial_{p_j}S_{p_i,p_i})\right],
\end{align*}
in which the right hand side only involves the derivatives of the $V$- and $S$-variables with respect to $p_j$,
and they can be further replaced by terms of $V$- and $S$-variables without derivatives with the help of \eqref{B:VDyn} and \eqref{B:VDyn} for special $a$ and $b$.
Finally, equations \eqref{B:tauDyna} and \eqref{B:tauDynb} help us to express every term on the right hand side of the above equation by the tau function.
As a result, we derive \eqref{ddBKP}.

We can also derive an analogue of \eqref{ddmAKP} in the BKP class, from the \ac{DL} scheme in this section.
Setting $a=p_i$ in \eqref{B:VDyn} yields
\begin{align*}
p_j\partial_{p_j}V_{p_i}=n_j\left[V_{p_i}\left(\frac{2p_ip_j}{p_i^2-p_j^2}-\frac{1}{2}V_{p_j}+\frac{1}{2}V_{-p_j}\right)+V_{p_j}\left(\frac{1}{2}\frac{p_j+p_i}{p_j-p_i}+S_{p_i,-p_j}\right)-V_{-p_j}\left(\frac{1}{2}\frac{p_j-p_i}{p_j+p_i}+S_{p_i,p_j}\right)\right].
\end{align*}
Substituting $V$ and $S$ according to \eqref{B:tauDyn}, we obtain \eqref{ddmBKP}.

An interesting observation is that the right sides of \eqref{ddBKP} and \eqref{ddmBKP} take the forms of
Hirota's discrete-time Toda equation (a 5-point equation) \cite{Hir2} and the bilinear discrete \ac{KdV} equation (a 6-point equation), respectively (see also chapter 8 of \cite{HJN16}),
though the parametrisation here is entirely different.

\section{Derivation of \eqref{ddCKP}}\label{S:CKP}

We choose the same plane wave factors as those of BKP in the CKP class, namely
\begin{align}\label{C:PWF}
\rho_{k}=\prod_{i=1}^\infty\left(\frac{p_i+k}{p_i-k}\right)^{n_i} \quad \hbox{and} \quad \sigma_{k'}=\rho_{k'}=\prod_{i=1}^\infty\left(\frac{p_i+k'}{p_i-k'}\right)^{n_i},
\end{align}
and these provide the same dynamical relations for the infinite matrix $\bC$ as follows:
\bse\label{C:CDyn}
\begin{align}
&(\rT_{n_j}\bC)\frac{p_j-\tbLd}{p_j+\tbLd}=\frac{p_j+\bLd}{p_j-\bLd}\bC, \label{C:CDyna} \\
&\partial_{p_j}\bC=n_j\left[\left(\frac{1}{p_j+\bLd}-\frac{1}{p_j-\bLd}\right)\bC+\bC\left(\frac{1}{p_j+\tbLd}-\frac{1}{p_j-\tbLd}\right)\right], \label{C:CDynb}
\end{align}
\ese
which describe the linear dispersion of the discrete CKP equation.

We select the Cauchy kernel
\begin{align}\label{C:Kernel}
\Oa_{k,k'}=\frac{1}{k+k'}
\end{align}
for the CKP class, which is the same as that in the AKP case, and therefore we still have the fundamental relation for $\bOa$ given by
\begin{align}\label{C:Omega}
\bOa\bLd+\tbLd\bOa=\bO.
\end{align}
Since the evolution of the infinite matrix $\bC$ is governed by \eqref{C:CDyn}, we reformulate \eqref{C:Omega} and present the following equations:
\bse\label{C:OmegaDyn}
\begin{align}
&\bOa\frac{p_j+\bLd}{p_j-\bLd}-\frac{p_j-\tbLd}{p_j+\tbLd}\bOa=2p_j\frac{1}{p_j+\tbLd}\bO\frac{1}{p_j-\bLd}, \label{C:OmegaDyna} \\
&\bOa\left(\frac{1}{p_j+\bLd}-\frac{1}{p_j-\bLd}\right)+\left(\frac{1}{p_j+\tbLd}-\frac{1}{p_j-\tbLd}\right)\bOa
=-\frac{1}{p_j-\tbLd}\bO\frac{1}{p_j+\bLd}-\frac{1}{p_j+\tbLd}\bO\frac{1}{p_j-\bLd}, \label{C:OmegaDynb}
\end{align}
\ese
in order to derive the dynamical relations of $\bU$ below.

Now we construct the dynamical evolutions of the infinite matrix $\bU$.
Acting the shift operator $\rT_{n_j}$ on equation \eqref{U} and taking \eqref{C:CDyna}, we have
\begin{align*}
\rT_{n_j}\bU=\left[1-(\rT_{n_j}\bU)\bOa\right](\rT_{n_j}\bC)=\left[1-(\rT_{n_j}\bU)\bOa\right]\frac{p_j+\bLd}{p_j-\bLd}\,\bC\,\frac{p_j+\tbLd}{p_j-\tbLd}.
\end{align*}
Notice that the infinite matrix $\bOa$ obeys \eqref{C:OmegaDyna}. We can rewrite the above equation as
\begin{align*}
(\rT_{n_j}\bU)\frac{p_j-\tbLd}{p_j+\tbLd}=\frac{p_j+\bLd}{p_j-\bLd}\bC-(\rT_{n_j}\bU)\bOa\frac{p_j+\bLd}{p_j-\bLd}\bC
=\frac{p_j+\bLd}{p_j-\bLd}\bC-(\rT_{n_j}\bU)\left(2p_j\frac{1}{p_j+\tbLd}\bO\frac{1}{p_j-\bLd}+\frac{p_j-\tbLd}{p_j+\tbLd}\bOa\right)\bC,
\end{align*}
and this can be further simplified as
\bse\label{C:UDyn}
\begin{align}\label{C:UDyna}
(\rT_{n_j}\bU)\frac{p_j-\tbLd}{p_j+\tbLd}=\frac{p_j+\bLd}{p_j-\bLd}\bU-2p_j(\rT_{n_j}\bU)\frac{1}{p_j+\tbLd}\bO\frac{1}{p_j-\bLd}\bU.
\end{align}
Equation \eqref{C:UDyna} is the first dynamical relation we need and it describes how $\bU$ evolves along the lattice directions $n_j$.
The other dynamical relation we need is the one evolving with regard to $p_j$, and the derivation is very similar to those of \eqref{A:UDynb} and \eqref{B:UDynb}, which reads
\begin{align}\label{C:UDynb}
\partial_{p_j}\bU={}&n_j\left[\left(\frac{1}{p_j+\bLd}-\frac{1}{p_j-\bLd}\right)\bU+\bU\left(\frac{1}{p_j+\tbLd}-\frac{1}{p_j-\tbLd}\right)\right.
\left.+\bU\frac{1}{p_j-\tbLd}\bO\frac{1}{p_j+\bLd}\bU+\bU\frac{1}{p_j+\tbLd}\bO\frac{1}{p_j-\bLd}\bU\right].
\end{align}
\ese

As is similar to the BKP case, we also need to impose a certain constraint on the spectral variables $k$ and $k'$.
This is realised by setting the integration domain $D$ symmetric and simultaneously requiring integration measure satisfying
\begin{align}\label{C:Reduction}
\rd\zeta(k',k)=\rd\zeta(k,k').
\end{align}
Such a reduction results in a symmetric infinite matrix $\bC$ because
\begin{align*}
\tbC=\iint_D\rd\zeta(k,k')\rho_k\bc_{k'}\tbc_k\rho_{k'}=\iint_D\rd\zeta(k',k)\rho_{k'}\bc_{k'}\tbc_k\rho_k=\bC.
\end{align*}
At the same time, we observe that the kernel \eqref{C:Kernel} is symmetric in terms of $k$ and $k'$, which implies that $\tbOa=\bOa$.
The symmetry properties of $\bC$ and $\bOa$ together guarantee that $\bU$ from \eqref{U} satisfies
\begin{align}\label{C:USym}
\tbU=\bU,
\end{align}
i.e. it is a symmetric infinite matrix.

During the derivation of \eqref{ddCKP} below, for convenience we introduce variables
\begin{align}
V_a\doteq 1-\left(\bU\frac{1}{a+\tbLd}\right)_{0,0}, \quad W_a\doteq 1-\left(\frac{1}{a+\bLd}\bU\right)_{0,0} \quad \hbox{and} \quad
S_{a,b}\doteq\left(\frac{1}{a+\bLd}\bU\frac{1}{b+\tbLd}\right)_{0,0},
\end{align}
which satisfy
\begin{align}
V_a=W_a \quad \hbox{and} \quad S_{a,b}=S_{b,a},
\end{align}
respectively, due to the symmetry condition \eqref{C:USym}.
We define the tau function in the CKP class as
\begin{align}\label{C:tau}
\tau=\det(1+\bOa\bC),
\end{align}
and after certain straightforward but relatively complex calculation based on \eqref{C:UDyna}, we can derive
\bse\label{C:tauDyn}
\begin{align}\label{C:tauDyna}
\frac{\rT_{n_j}\tau}{\tau}=1+2p_jS_{-p_j,-p_j} \quad \hbox{and} \quad \frac{\rT_{n_j}^{-1}\tau}{\tau}=1-2p_jS_{p_j,p_j},
\end{align}
as well as
\begin{align}\label{C:tauDynb}
\left[1-(p_i+p_j)S_{p_i,p_j}\right]^2=\frac{(p_i+p_j)^2(\rT_{n_i}^{-1}\tau)(\rT_{n_j}^{-1}\tau)-(p_i-p_j)^2\tau(\rT_{n_i}^{-1}\rT_{n_j}^{-1}\tau)}{4p_ip_j\tau^2}.
\end{align}

Similar to the BKP case, the $S_{-p_i,p_j}$, $S_{p_i,-p_j}$ and $S_{-p_i,-p_j}$ analogues of \eqref{C:tauDynb}
follow from the interchange relation $(p_i,\rT_{n_i})\leftrightarrow(-p_i,\rT_{n_j}^{-1}$.
Moreover, the dynamics of the tau function with respect to the lattice parameter $p_j$ is derived by carrying the same calculation in the derivation of \eqref{A:tauDynb},
which establishes the link between $\tau$ and $S$ and takes the form of
\begin{align}
\partial_{p_j}\ln\tau=2n_j S_{p_j,-p_j}.
\end{align}
\ese

Notice from \eqref{C:tauDyn} that the tau function is connected with the $S$-variable.
We therefore multiply \eqref{C:UDynb} by $(a+\bLd)^{-1}$ from the left and $(b+\tbLd)^{-1}$ from the right simultaneously and take the $(0,0)$-entry,
and as a result, an equation for $S_{a,b}$ arises in the form of
\begin{align*}
\partial_{p_j}S_{a,b}={}&n_j\left[\frac{1}{p_j-a}(S_{a,b}-S_{p_j,b})+\frac{1}{p_j+a}(S_{-p_j,b}-S_{a,b})\right. \nonumber \\
&\qquad\left.+\frac{1}{b-p_j}(S_{a,p_j}-S_{a,b})+\frac{1}{p_j+b}(S_{a,-p_j}-S_{a,b})-S_{a,-p_j}S_{p_j,b}-S_{a,p_j}S_{-p_j,b}\right].
\end{align*}
By setting $a=p_i$ and $b=-p_i$, it is rewritten as
\begin{align*}
\partial_{p_j}S_{p_i,-p_i}={}&\frac{n_j}{(p_i+p_j)^2}\left[(1-(p_i+p_j)S_{p_i,p_j})(1+(p_i+p_j)S_{-p_i,-p_j})-1\right] \nonumber \\
&\qquad+\frac{n_j}{(p_i-p_j)^2}\left[(1-(p_i-p_j)S_{p_i,-p_j})(1-(p_j-p_i)S_{-p_i,p_j})-1\right].
\end{align*}
It is then possible to replace the $S$-variable by $\tau$ through \eqref{C:tauDyna} and \eqref{C:tauDynb} in the above equation.
As a result, it leads to the nonautonomous differential-difference CKP equation \eqref{ddCKP}.
We note that this equation is still in the form of the (2+2)-dimensional Toda-type,
but here the bilinearity is broken (though the equation is still quadratic) compared with \eqref{ddAKP} and \eqref{ddBKP} since the square root operation is involved.
This is not surprising as we have seen the discrete CKP equation \eqref{dCKP} is a quartic equation.
Equation \eqref{ddCKP} can alternatively be written as an eighth-power equation
\begin{align*}
\left\{\left[\frac{p_ip_j(p_i^2-p_j^2)^2}{n_in_j}\rD_{p_i}\rD_{p_j}\tau\cdot\tau
+8p_ip_j(p_i^2+p_j^2)\tau^2\right]^2-(p_i-p_j)^4EF-(p_i+p_j)^4GH\right\}^2=4(p_i^2-p_j^2)^4EFGH
\end{align*}
by taking square twice in order to eliminate the square root in the expression, where $E$, $F$, $G$ and $G$ are already given in section \ref{S:Intro} below \eqref{dCKP}.

\section{Remarks on soliton solutions, multi-dimensional consistency and higher-order symmetries}\label{S:Integr}

We have shown how the six nonautonomous \ac{DDeltaE}s are derived from the \ac{DL} scheme,
by associating each equation with a certain linear integral equation in the form of \eqref{Integral}.
From the viewpoint of the \ac{DL} approach, this guarantees the integrability of the resulting nonlinear \ac{DDeltaE}s,
as long as their respective linear integral equations are solvable.
Here by integrability we mean the existence of an exact solution possessing an infinite number of free parameters (e.g. soliton-type solution),
while the initial boundary value problem is not discussed.
Below we present soliton solutions of these \ac{DDeltaE}s, by performing a very special reduction on the measure in the respective linear integral equations.

We start with the formal tau functions \eqref{A:tau}, \eqref{B:tau} and \eqref{C:tau} in order to construct soliton-type solutions,
instead of the linear integral equation \eqref{Integral},
since the tau functions defined as such already contain the key ingredients of the linear integral equation,
namely the integration domain and measure, the plane wave factors as well as the Cauchy kernel.
The procedure is exactly the same as that given in \cite{Fu17a}. Hence, we only provide a short guide and directly list the results below.
For equations \eqref{ddAKP}, \eqref{ddmAKP} and \eqref{ddumAKP}, we take the measure
\begin{align}\label{A:Singular}
\rd\zeta(k,k')=\sum_{i=1}^{N}\sum_{j=1}^{N'}\frac{A_{i,j}}{(2\pi\mathfrak{i})^2}\frac{1}{k-k_i}\frac{1}{k'-k'_j}\rd k\rd k' \quad
\hbox{with $\mathfrak{i}$ being the imaginary unit},
\end{align}
which results in the determinant solution (i.e. soliton solution)
\begin{align}\label{A:Sol}
\tau=\det(1+\bA\bM),
\end{align}
where $1$ denotes the $N\times N$ unit matrix, $\bA=(A_{i,j})_{N\times N'}$ is an arbitrary matrix and the entries of $\bM=(M_{j,i})_{N'\times N}$ are given by
\begin{align*}
M_{j,i}=\frac{\rho_{k_i}\sigma_{k'_j}}{k_i+k'_j}, \quad \hbox{with $\rho$ and $\sigma$ given by \eqref{A:PWF}}.
\end{align*}
Here $N$ and $N'$ are arbitrary positive integers.
Similarly, for equations \eqref{ddBKP} and \eqref{ddmBKP} we take a special measure
\begin{align}\label{B:Singular}
\rd\zeta(k,k')=\sum_{i,j=1}^{2N}\frac{A_{i,j}}{(2\pi\mathfrak{i})^2}\frac{1}{k-k_i}\frac{1}{k'-k'_j}\rd k\rd k', \quad A_{i,j}=-A_{j,i},
\end{align}
which obeys the antisymmetry condition in \eqref{B:Reduction}.
In this case, the tau function is determined by
\begin{align}\label{B:Sol}
\tau^2=\det(1+\bA\bM),
\end{align}
where $\bA=(A_{i,j})_{2N\times2N}$ is an antisymmetric matrix and the Cauchy matrix $\bM=(M_{j,i})_{2N\times2N}$ has its $(j,i)$-entry
\begin{align*}
M_{j,i}=\rho_{k_i}\frac{1}{2}\frac{k_i-k'_j}{k_i+k'_j}\sigma_{k'_j}, \quad \hbox{in which $\rho$ and $\sigma$ are defined by \eqref{B:PWF}}.
\end{align*}
For the CKP class, we consider the following measure reduction:
\begin{align}\label{C:Singular}
\rd\zeta(k,k')=\sum_{i,j=1}^{N}\frac{A_{i,j}}{(2\pi\mathfrak{i})^2}\frac{1}{k-k_i}\frac{1}{k'-k'_j}\rd k\rd k', \quad A_{i,j}=A_{j,i},
\end{align}
where we have respected the condition in \eqref{C:Reduction}.
Then the formal tau function \eqref{C:tau} turns out to be
\begin{align}\label{C:Sol}
\tau=\det(1+\bA\bM),
\end{align}
in which $\bA=(A_{i,j})_{N\times N}$ is a symmetric matrix and the $(i,j$)-entry of $\bM=(M_{j,i})_{N,N}$ is defined as
\begin{align*}
M_{j,i}=\frac{\rho_{k_i}\sigma_{k'_j}}{k_i+k'_j}, \quad \hbox{where $\rho$ and $\sigma$ are defined as \eqref{C:PWF}}.
\end{align*}

The \ac{DL} also provides a perspective to understand the \ac{MDC} property of the six \ac{DDeltaE}s, from their respective solution structures.
We focus on the formal tau functions \eqref{A:tau}, \eqref{B:tau} and \eqref{C:tau},
since the soliton-type expressions \eqref{A:Sol}, \eqref{B:Sol} and \eqref{C:Sol} are the natural consequences
arising from the measure reductions \eqref{A:Singular}, \eqref{B:Singular} and \eqref{C:Singular}, respectively.
We take equation \eqref{ddBKP} as an example.
The formal solution \eqref{B:tau} relies on its flow variables $p_i$, $p_j$, $n_i$ and $n_j$ through the infinite matrix $\bC$,
whose dynamics are completely determined by the plane wave factor $\rho_k\sigma_{k'}$ given in \eqref{B:PWF}.
Observing that all independent variables are on the same footing\footnote{
This is also the reason why we adopt the notations $p_j$ and $n_j$ for $j=1,2,\cdots$ rather than select fixed indices from the beginning.
} in \eqref{B:PWF}, we conclude that \ac{DDeltaE}s in the form of \eqref{ddBKP} with regard to different lattice parameters and variables
share the same nontrivial solution \eqref{B:tau}; in other words, these equations are compatible with each other.
This is the \ac{MDC} property that we normally adopt in the theory of discrete integrable systems, and here it is applied to the differential-difference case.

Furthermore, the derivation of \eqref{ddAKP}, \eqref{ddmAKP} and \eqref{ddumAKP} shows that the three equations have the same tau function \eqref{A:tau} as the common solution,
and such a tau function simultaneously (cf. \cite{Fu17a}) serves a solution of \eqref{dAKP}.
This implies that these four equations are multi-dimensionally consistent, within the \ac{DL} scheme;
in other words, equations \eqref{ddAKP}, \eqref{ddmAKP}, \eqref{ddumAKP} and \eqref{dAKP} should be treated as compatible flows in the AKP class.
Similarly, equations \eqref{ddBKP}, \eqref{ddmBKP} and \eqref{dBKP} are also multi-dimensionally consistent with each other;
while equation \eqref{ddCKP} is compatible with \eqref{dCKP}, from the same viewpoint.
This is also the \ac{MDC} property, namely equations in different forms can still be consistent with each other.
As a remark, we would like to note that here the \ac{MDC} is only verified on the solutions generated by the \ac{DL} framework.
The verification of the consistency on the level of equations still remains future work.
We believe the assertion is correct, as the solutions following from the \ac{DL} already possess an infinite number of degrees of freedom.

With the help of the solution structures of the AKP, BKP and CKP classes discussed above,
we are now allowed to explain how the higher-order symmetries in the continuous \ac{KP}-type hierarchies are generated from the lattice parameters.
The idea to realise this is to compare the effective plane wave factors in each discrete \ac{KP} class with its corresponding continuous analogues (which are given in \cite{Fu18b}).  In the AKP class, we require
\begin{align*}
\rho_k\sigma_{k'}=\prod_{i=1}^\infty\left(\frac{p_i+k}{p_i-k'}\right)^{n_i}
=\exp\left\{\sum_{j=1}^\infty\left(k^j-(-k')^j\right)\frac{(-1)^{j-1}}{j}\sum_{i=1}^\infty\frac{n_i}{p_i^j}\right\}
=\exp\left\{\sum_{j=1}^\infty\left(k^j-(-k')^j\right)x_j\right\}.
\end{align*}
Therefore, the relationship between the continuous variables $x_j$ and the lattice parameters $p_i$ is given by
\begin{align}\label{A:Sym}
x_j=\frac{(-1)^{j-1}}{j}\sum_{i=1}^\infty\frac{n_i}{p_i^j} \quad \hbox{for} \quad j=1,2,\cdots, \quad \hbox{which results in} \quad \partial_{p_i}=\sum_{j=1}^\infty(-1)^j\frac{n_i}{p_i^{j+1}}\partial_{x_j}.
\end{align}
Likewise, as the form of the plane wave factor is different, we require in the BKP and CKP classes
\begin{align*}
\rho_k\sigma_{k'}=\prod_{i=1}^\infty\left(\frac{p_i+k}{p_i-k}\frac{p_i+k'}{p_i-k'}\right)^{n_i}
=\exp\left\{\sum_{j=0}^\infty\left(k^{2j+1}+k'^{2j+1}\right)\frac{2}{2j+1}\sum_{i=1}^\infty\frac{n_i}{p_i^{2j+1}}\right\}
=\exp\left\{\sum_{j=0}^\infty\left(k^{2j+1}+k'^{2j+1}\right)x_{2j+1}\right\},
\end{align*}
and subsequently we obtain
\begin{align}\label{BC:Sym}
x_{2j+1}=\frac{2}{2j+1}\sum_{i=1}^\infty\frac{n_i}{p_i^{2j+1}} \quad \hbox{for} \quad j=0,1,2,\cdots, \quad \hbox{which leads to} \quad \partial_{p_i}=-2\sum_{j=0}^\infty\frac{n_i}{p_i^{2j+2}}\partial_{x_{2j+1}}.
\end{align}
The transformations between the derivatives of lattice parameters $p_i$ and continuous flow variables $x_j$ listed in \eqref{A:Sym} and \eqref{BC:Sym}
provide a way of generating all the higher-order symmetries of the continuous \ac{KP}-type from the lattice parameters.
They are the counterparts of the result in the generating \ac{PDE} of the discrete \ac{KdV} equation given in \cite{NHJ00} on the \ac{KP} level,
though the generating \ac{PDE}s for the discrete KP-type equations are not yet clear so far.

\section{Conclusions}\label{S:Concl}

Based on the \ac{DL} framework for the discrete AKP, BKP and CKP equations,
we introduce a new perspective to construct integrable nonlinear equations,
by thinking of the lattice parameters as the independent variables
and treating them together with the discrete variables equally.
As a result, six novel nonautonomous differential-equations are proposed,
including two (2+2)-dimensional nonautonomous \ac{DDeltaE}s, namely \eqref{ddBKP} and \eqref{ddCKP} from BKP and CKP classes, respectively,
as well as four (2+1)-dimensional nonautonomous semi-discrete equations,
i.e. \eqref{ddAKP}, \eqref{ddmAKP} and \eqref{ddumAKP} from the AKP class and \eqref{ddmBKP} from the BKP class.

It was already shown that nonautonomous \ac{DDeltaE}s often play the role of master symmetries
for lower-dimensional discrete integrable systems \cite{TTP05},
while the master symmetry theory for higher-dimensional lattice equations is not yet clear.
These new equations potentially provide us with an insight into understanding master symmetries of \ac{3D} integrable discrete models.
Furthermore, it seems that (2+2)-dimensional nonautonomous integrable models of Toda-type, such as equations \eqref{ddBKP} and \eqref{ddCKP},
have never appeared in the literature so far, to be best of the authors' knowledge.
Their geometric interpretation still needs to be discovered.

In addition to symmetries and master symmetries, we also note that the integrability of discrete \ac{KP}-type equations was also proven from the perspective of conservations laws.
In \cite{MQ06}, conservation laws for the discrete AKP and BKP equations were constructed, while it remains a problem for the discrete CKP equation.

The ultimate goal is to find closed-form generating \ac{PDE}s for higher-dimensional integrable hierarchies and their integrability structures.
This also remains an open problem for future study.

\section*{Acknowledgments}
WF was sponsored by the National Natural Science Foundation of China (grant nos. 11901198 and 11871396) and by Shanghai Pujiang Program (grant no. 19PJ1403200).
This project was also partially supported by the Science and Technology Commission of Shanghai Municipality (grant no. 18dz2271000).

\renewcommand{\bibname}{References}
\bibliography{References}
\bibliographystyle{amsplain}

\end{document}